\documentclass[9pt,twocolumn,twoside]{pnas-new}
\templatetype{pnasresearcharticle} % Choose template
\usepackage{graphicx}

\usepackage[T1]{fontenc}
\usepackage[utf8]{inputenc}
\usepackage{amsmath}
\usepackage{color}
\usepackage{mathtools}
\usepackage{amssymb}
\usepackage{hyperref}
\usepackage{natbib}
\usepackage{subcaption}

\newcommand{\dd}[1]{\mathrm{d}#1 \,}

\newcommand{\dev}[2]{\frac{\mathrm{d}#1}{\mathrm{d}#2}}
\renewcommand{\v}[1]{\boldsymbol{#1}}

\newcommand\odequ{\mathrel{\stackrel{\makebox[0pt]{\mbox{\normalfont\tiny 1D}}}{=}}}

\begin{document}

\title{Relaxation to universal non-Maxwellian equilibria in a collisionless plasma}

% Use letters for affiliations, numbers to show equal authorship (if applicable) and to indicate the corresponding author
\author[a,b,1]{R. J. Ewart}
\author[a,c]{M. L. Nastac}
\author[d]{P. J. Bilbao}
\author[d]{T. Silva}
\author[d]{L. O. Silva}
\author[a,e]{A. A. Schekochihin}

\affil[a]{Rudolf Peierls Centre for Theoretical Physics, University of Oxford, Oxford, OX1 3PU, UK}
\affil[b]{Balliol College, Oxford, OX1 3BJ}
\affil[c]{St John's College, Oxford, OX1 3JP}
\affil[d]{GoLP/Instituto de Plasmas e Fus\~{a}o Nuclear, Instituto Superior T\'{e}cnico, Universidade de Lisboa, 1049-001 Lisbon, Portugal}
\affil[e]{Merton College, Oxford, OX1 4JD}

% Please give the surname of the lead author for the running footer
\leadauthor{Ewart}

% Please add a significance statement to explain the relevance of your work
\significancestatement{A fundamental tenet of thermodynamics is that chaotic systems will relax to maximum-entropy states. In plasmas, the chaos is conventionally provided by interparticle collisions and the universal maximum-entropy equilibrium is a Maxwellian distribution. However, in collisionless plasmas, the chaotic state is due to collective turbulent dynamics. In this paper, we argue theoretically, and show numerically, that such plasmas still relax towards universal equilibria, but are non-Maxwellian, featuring a power-law distribution of particle energies. These equilibria are still obtained via a maximum-entropy principle, but taking into account the conservation, on short time scales, of phase volume in a collisionless system, and a subtle form of breaking of this conservation on longer time scales, due to the nature of collisionless plasma turbulence.}

% Please include corresponding author, author contribution and author declaration information
\authorcontributions{R.J.E., L.O.S., and A.A.S. designed research; R.J.E., M.L.N., P.J.B., T.S., L.O.S, and A.A.S. contributed new reagents/analytic tools; R.J.E., M.L.N., P.J.B, and T.S. analyzed data; R.J.E., M.L.N., P.J.B., T.S., L.O.S., and A.A.S. wrote the paper.}
\authordeclaration{The authors have no competing interests to declare.}%The authors have no competing interests to declare
\correspondingauthor{\textsuperscript{1}To whom correspondence should be addressed. E-mail: robert.ewart@physics.ox.ac.uk}

% At least three keywords are required at submission. Please provide three to five keywords, separated by the pipe symbol.
\keywords{plasma $|$ statistical mechanics $|$ non-thermal particle acceleration}

\begin{abstract}
Generic equilibria are derived for turbulent relaxing plasmas via an entropy-maximization procedure that accounts for the short-time conservation of certain collisionless invariants. The conservation of these collisionless invariants endows the system with a partial `memory' of its prior conditions, but is imperfect on long time scales due to the development of a turbulent cascade to small scales, which breaks the precise conservation of phase volume, making this memory imprecise. The equilibria are still determined by the short-time collisionless invariants, but the invariants themselves are driven to a universal form by the nature of the turbulence. This is numerically confirmed for the case of beam instabilities in one-dimensional electrostatic plasmas, where sufficiently strong turbulence appears to cause the distribution function of particle energies to develop a universal power-law tail, with exponent $-2$. 
\end{abstract}

\dates{This manuscript was compiled on June 31, 2024}
\doi{\url{www.pnas.org/cgi/doi/10.1073/pnas.XXXXXXXXXX}}

\maketitle
\thispagestyle{firststyle}
\ifthenelse{\boolean{shortarticle}}{\ifthenelse{\boolean{singlecolumn}}{\abscontentformatted}{\abscontent}}{}

\firstpage[23]{2}
\dropcap{T}he naive application of statistical mechanics would imply that one should find a universe full to the brim with plasma in local Maxwell--Boltzmann equilibrium. In fact, the plasmas in our Universe are not, by and large, in such an equilibrium. This is true in many settings, from cosmic rays populating our Galaxy and beyond~\cite{Becker_2020}, down through the solar neighbourhood~\cite{Oka_2018,Wilson_2022}, and all the way to Earth-based experiments~\cite{Cruz_2018,Magee_2019,Hartouni_2022}. It is not at all difficult, of course, to motivate why this should be the case. The process by which plasmas are driven towards Maxwell--Boltzmann equilibrium---two-body interparticle collisions~{\cite{Boltzmann_1896,Landau_1936,Balescu_1960,Lenard_1960}}---typically takes place over a time scale much longer than that associated with the evolution set by the mean (averaged over length scales far larger than the interparticle separation) electric and magnetic fields that the plasma inherently generates. In systems with large-scale (system-size) gradients, these fields may be driven unstable, triggering not just a zoo but an entire ecosystem of plasma instabilities \cite{Krall_1973, Bott_2024}, with fluctuations growing and reaching amplitudes at which they can react back upon their progenitors, potentially altering the mean distribution and relaxing the system towards some semblance of stability, but by no means Maxwellianity. Despite all this, the distributions that we observe do possess some universal features (or fall into a finite number of universality classes \cite{Marsch_2006,Pierrard_2010,Zhdankin_2017, Verscharen_2019}).

It has therefore been a question of interest since the earliest observations of non-thermal distributions whether they can be explained by an overarching theory of relaxation. One such theory, proposed  by Lynden-Bell \cite{LyndenBell_1967} in the context of self-gravitating systems, gave hope of an affirmative answer to this question by appealing to the statistical-mechanical principle of maximum entropy even in the absence of particle collisions to enforce it. Unlike the collisional approach, Lynden-Bell's theory made use of the fact that the distribution function $f(\v{x},\v{v})$ of a collisionless system naturally obeys a Liouville equation and, therefore, conserves an infinite family of invariants: the conservation of `phase volume' in position and velocity space~$(\v{x},\v{v})$ implies that, for any function $G(f)$, 
\begin{equation}
\label{eqn:E1}
\int \dd{\v{x}}\dd{\v{v}} G(f)  = \mathrm{const.}
\end{equation}
Therefore, in line with the dictates of statistical mechanics, the conservation of this continuum of quantities, known as `Casimir invariants'~\cite{Ye_1992}, must be respected when the entropy of the system (appropriately defined) is maximized---effectively giving the system an enhanced memory of its initial conditions. Despite the promise of this theory, its actual application has been limited due to the inherent complexity and apparent non-universality of the resulting equilibria~\cite{Arad_2005,Arad_2005b,Levin_2008,Assllani_2012,Levin_2014}. A conceptually graver concern is the reliance on the \textit{precise} conservation of the Casimir invariants. In any real system, their conservation was only ever going to be approximate, with molecular chaos eventually reasserting itself. Aggravating this concern further, it has been demonstrated both theoretically~{\cite{Schekochihin_2008,Eyink_2018,Nastac_2024}} and numerically~\cite{Tatsuno_2009, Zhdankin_2021a} that even for systems with nominally weak interparticle collisions---such that the time scale for relaxation to a Maxwellian equilibrium is much longer than dynamical times---the Casimir invariants are broken on time scales competitive with the evolution of the system (provided there is a sufficient level of turbulence to stir it), rendering their status as invariants questionable at best.

In this paper, we  show that turbulent plasmas do achieve Lynden-Bell equilibria, and that, rather than doing this in spite of the breaking of Casimir invariants, they manage it \textit{in tandem with} this. We further show that the Lynden-Bell equilibrium that is achieved has a universal high-energy asymptotic. 
\begin{figure*}[t!]
\centering
\includegraphics[width=17.8cm,height=10.16cm]{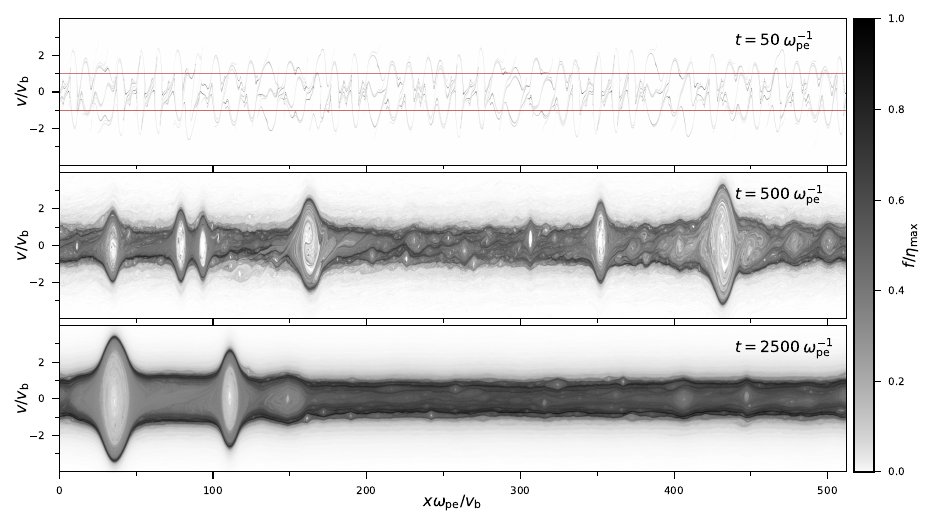}
\caption{\label{Fig:1}Self-consistent particle-in-cell simulation of the temporal evolution (from top to bottom) of the phase-space density~$f(x,v)$ for the electron-only two-stream instability, visualised across three time snapshots:~$50\, \omega_{\mathrm{pe}}^{-1}$,~$500\, \omega_{\mathrm{pe}}^{-1}$, and~$2500\, \omega_{\mathrm{pe}}^{-1}$ (where $\omega_{\mathrm{pe}}^{-1}$ is the electron plasma frequency). The two counter-propagating beams, of speed~$v_{\mathrm{b}}$, (top in red) rapidly go unstable generating the phase-space structure shown at~$50\, \omega_{\mathrm{pe}}^{-1}$. As the system evolves (middle and bottom snapshots), the formation and merger of turbulent electron holes is clearly evident, which drives the mixing of the phase space and relaxation towards a collisionless equilibrium. The color scale is normalised to the peak value of~$f(x,v)$ in each snapshot.}
\end{figure*}
We carry out this study the aid of the particle-in-cell (PIC) code OSIRIS~\cite{Fonseca_2002}, applied to one of the most classically studied turbulent collisionless systems: the two-stream instability. The very earliest simulations~\cite{Roberts_1967,Morse_1969} showed vividly that this instability leads a system composed of long-lived structures in phase space (cf. `BGK modes'~\cite{Bernstein_1957}), known as  electron `holes' (see figure~\ref{Fig:1}). These holes, since studied analytically, numerically, and observationally (see, e.g.,~\cite{Hutchinson_2017,Hutchinson_2024} and references therein) move around, merge, and generically represent a relaxing turbulent state. It is precisely this turbulence that drives the system's relaxation towards the Lynden-Bell equilibrium. However, the same turbulence generates small velocity- and position-space structures (in the manner of a turbulent cascade predicted by~\cite{Nastac_2024b}), which cause the Casimir invariants to evolve with time. In the course of this evolution, the underlying Lynden-Bell equilibrium to which the system wants to relax is gradually changed and the system adjusts to reach this evolving target equilibrium. Thus, the system's precise memory of initial conditions is replaced with a `turbulent amnesia': turbulent fluctuations are perpetually trying to push the system towards its Lynden-Bell equilibrium, but the goalposts are continually moved by the breaking of Casimir invariants driven by those same fluctuations. Eventually the Casimir invariants themselves reach a steady state, previously conjectured by~\cite{Ewart_2023}, which causes the system to converge to a final, universal Lynden-Bell equilibrium. This equilibrium exhibits a particle-energy distribution possessing a power-law tail with exponent $-2$, confirming the existence of universal equilibria in strongly turbulent relaxing plasmas.
\section*{Lynden-Bell equilibria}
\begin{figure*}[t!]
\centering
\includegraphics[width=17.8cm,height=8.89cm]{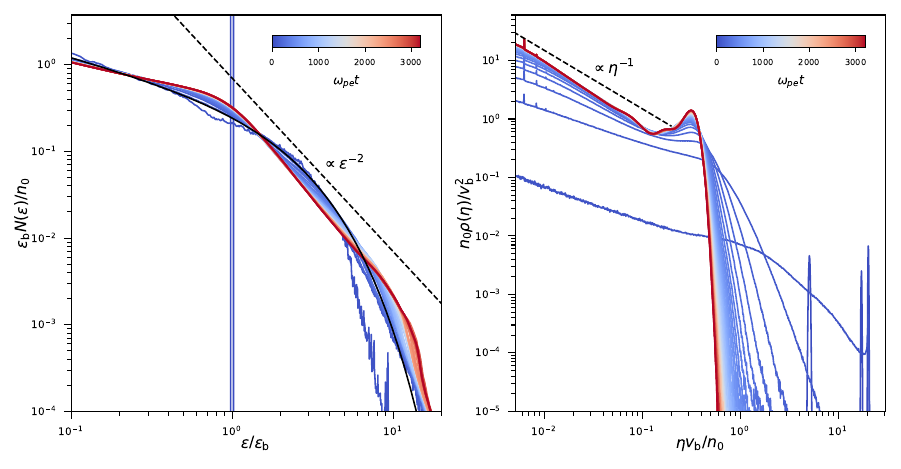}
\caption{\label{Fig:2}Left panel: the evolution of the mean distribution function $N(\varepsilon)$ of particle energies $\varepsilon = mv^{2}/2$ for the electron two-stream instability. The energies are normalized to the initial beam-particle energy~$\varepsilon_{\mathrm{b}} = mv_{\mathrm{b}}^{2}/2$;~$n_{0}$ is the mean particle number density.  The solid black line shows a Maxwellian of the same energy as the initial condition. The lines shown in shades of the blue-to-red color palette are the distributions from the initial condition (dark blue) to $t=3200\, \omega_{\mathrm{pe}}^{-1}$ (dark red) at successive time intervals of~$50 \, \omega_{\mathrm{pe}}^{-1}$. Right panel: the evolution of the `waterbag content' $\rho(\eta)$, over time, computed from [\ref{eqn:E7}] for the fine-grained phase-space density $f(\v{x},\v{v})$, as detailed in the Materials and Methods. The color scheme for time evolution is the same as in the left panel. The universal asymptotics of $N(\varepsilon)$ and $\rho(\eta)$ are plotted as dashed lines showing the development of a power-law tail~[\ref{eqn:E17}] and the agreement with the predicted scaling of the waterbag content~[\ref{eqn:E15}].}
\end{figure*} 
The formalism of Lynden-Bell statistical mechanics is built around the argument that, during relaxation, the plasma becomes highly chaotic and disordered in phase space. A prototypical example of this, a plasma experiencing a two-stream instability (the simulation and statistical mechanics of which will be discussed below), is shown in figure~\ref{Fig:1}: the plasma starts as two cold counter-propagating electron beams in phase space and is mixed until a myopic eye cannot distinguish the forest (mean phase-space density) from the trees (exact phase-space density). Because of this disorder, and in light of the conservation of Casimir invariants [\ref{eqn:E1}], it is helpful to consider the exact phase-space density $f(\v{x},\v{v})$ to be a random quantity and to describe the statistical state of the plasma by the probability density~$P(\v{v},\eta)$ (which can be defined over a large spatial average) that, at a given velocity~$\v{v}$, this phase-space density~$f(\v{x},\v{v})$ takes the value~$\eta$ \cite{Chavanis_1996,Ewart_2022}. Once~$P(\v{v},\eta)$ is known, the mean phase-space density~$\langle f\rangle$ can always be computed as the first moment in $\eta$ of $P(\v{v},\eta)$:
\begin{equation}
\label{eqn:E2}
\langle f\rangle(\v{v}) = \int\dd{\eta}\eta P(\v{v},\eta).
\end{equation} 
To determine $P(\v{v},\eta)$, in equilibrium, one assumes that the system maximizes the Lynden-Bell entropy
\begin{equation}
\label{eqn:E3}
S_{\mathrm{LB}} = -\int\dd{\v{v}}\dd{\eta}P(\v{v},\eta)\ln P(\v{v},\eta)
\end{equation}
subject to reasonable constraints\footnote{The choice to maximize entropy~$S_{\mathrm{LB}}$, while natural from an information-theoretic standpoint, is not free of assumptions. A crucial one is that the values of the exact phase-space density~$f(\v{x},\v{v})$ at different~$\v{x}$ and~$\v{v}$ in a given dynamical state (provided that state satisfies the required constraints) are independent at some fine-grained scale. This could be thought of as a hypothesis of perfect mixing---we shall discuss its validity later on.}. The most obvious such constraint is the conservation of probability:
\begin{equation}
\label{eqn:E4}
\int \dd{\eta} P(\v{v},\eta) = 1.
\end{equation}
The next most unobjectionable constraint is that the total energy of the system is fixed:
\begin{equation}
\label{eqn:E5}
\int\dd{\v{v}}\frac{1}{2}m|\v{v}|^{2}\int \dd{\eta}\eta P(\v{v},\eta) = E_{\mathrm{tot}},
\end{equation}
where we have neglected contributions from the potential energy (which we will find to be smaller by a factor~$\sim\!100$ in our simulations). The final constraint that must be enforced is the conservation of phase volume, mathematically stated as 
\begin{equation}
\label{eqn:E6}
\int\dd{\v{v}}P(\v{v},\eta) = \rho(\eta).
\end{equation}
This simply reads that the volume-integrated probability of the exact phase-space density~$f$ taking the value~$\eta$ is equal to some prescribed function of~$\eta$. This function, referred to by us as the `waterbag content', can be computed from the exact phase-space density of the plasma:
\begin{equation}
\label{eqn:E7}
\rho(\eta) = \frac{1}{V}\iint\dd{\v{x}}\dd{\v{v}}\delta (f(\v{x},\v{v}) - \eta),
\end{equation}
where $V$ is the spatial volume of the system. In effect, $\rho(\eta)$ measures the phase-space volume occupied by each level set~$\eta$ of~$f$. This indeed captures all Casimir invariants~[\ref{eqn:E1}], which can be recovered as weighted integrals of~$\rho(\eta)$\footnote{While we have accounted for the infinite set of Casimir invariants~[\ref{eqn:E1}], those do not include every possible ideal invariant. For instance, the Vlasov equation also preserves topological invariants, such as the connectivity of the level sets of the phase-space density~\cite{LyndenBell_1967}.}:
\begin{equation}
\label{eqn:E8}
\int \dd{\v{x}}\dd{\v{v}} G(f) = V\int \dd{\eta} G(\eta)\rho(\eta).
\end{equation}

Thus, $P(\v{v},\eta)$ naturally enables the book-keeping of the level sets of the phase-space density. There is a subtle complication (discussed in greater detail in~\cite{Chavanis_2006b,Ewart_2023}) as to how~$\eta = 0$ (the empty level set) is handled, since $\rho(\eta)$ cannot be finite at~$\eta = 0$ for systems with unbounded velocity domains. Treating this subtlety carefully and maximising the entropy~[\ref{eqn:E3}] subject to the constraints~[\ref{eqn:E4}-\ref{eqn:E6}] gives the Lynden-Bell equilibria
\begin{equation}
\label{eqn:E9}
P(\v{v},\eta) = \frac{\delta(\eta) + e^{-\beta \eta \varepsilon(\v{v})}F(\eta)}{1+ \int_{\eta>0}\dd{\eta'} e^{-\beta \eta' \varepsilon(\v{v})}F(\eta')},
\end{equation}
where $\varepsilon(\v{v}) = m|\v{v}|^{2}/2$ and $\beta$ and $F(\eta)$ are Lagrange multipliers, known, in analogy with conventional thermodynamics, as the inverse `thermodynamic temperature' and the `fugacity', respectively. They are specified by enforcing the constraints~[\ref{eqn:E5}] and~[\ref{eqn:E6}] for the function~[\ref{eqn:E9}] (which has already been normalized to satisfy [\ref{eqn:E4}]). 
\subsection*{`Turbulent amnesia'}
The formalism of Lynden-Bell is, therefore, now completely prescriptive: in specifying the initial condition, one chose the energy of the system~$E_{\mathrm{tot}}$ and the waterbag content~$\rho(\eta)$ through~[\ref{eqn:E7}]. Strict adherence to Lynden-Bell's theory would then imply that the system should relax towards the appropriately solved equilibrium~[\ref{eqn:E9}]. This is however, not the case. In any real situation, including any numerical simulation, one will find that the mean phase-space density will continue evolving even after the initial instability is quenched---and the putative collisionless invariants~$\rho(\eta)$ will evolve as well. This is manifest in figure~\ref{Fig:2}, showing the energy distribution~$N(\varepsilon)$ of particles, and their waterbag content~$\rho(\eta)$ for a plasma undergoing the nearly collisionless two-stream instability depicted in figure~\ref{Fig:1}. It is, of course, clear why~$\rho(\eta)$ should change: no truly collisionless plasma (or simulation) can exist, the Liouville equation is never perfectly satisfied, and so the memory of the initial conditions encoded by~$\rho(\eta)$ cannot be preserved forever. Why~$\rho(\eta)$ evolves relatively fast, even for nearly collisionless systems, will be discussed below. For the purposes of formulating a theoretical scenario and verifying it numerically, it suffices to know that~$\rho(\eta)$ does evolve. We therefore propose an amendment to the Lynden-Bell theory: the principle of `turbulent amnesia'. Under this scheme, the collisionless dynamics of the plasma push the system towards the Lynden-Bell equilibrium computed using the time-evolving~$\rho(\eta)$, which is changed as the turbulence scrambles the system's long-term memory of its prior conditions. This is somewhat analogous to the way in which collisional plasmas undergoing heating will pass through a sequence of Maxwellian distributions with distinct temperatures even though the energy is not a conserved quantity: systems still strive to maximize entropy rapidly even when their invariants are imperfect. We shall return to the theoretical justification of this scenario once we have established its validity numerically. 
\section*{Numerical verification}
\begin{figure}[t!]
\centering
\includegraphics[width=8.68cm,height=8.68cm]{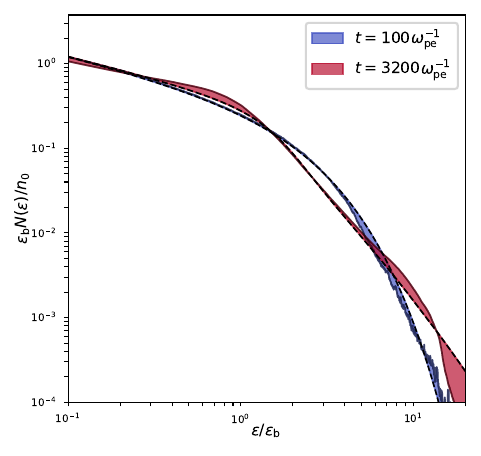}
\caption{\label{Fig:3}Comparison of the directly measured mean phase-space density of the system (solid lines) and the Lynden-Bell equilibrium obtained by taking the first $\eta$ moment [\ref{eqn:E2}] of [\ref{eqn:E9}] (dashed black lines) at two different times. At early times (blue), the mean phase-space density closely follows the Maxwellian associated with $\rho(\eta)$ computed from the initial condition (the two beams, which have a single value of~$\eta$), while at late times (red), it is better fit by the Lynden-Bell equilibrium associated with the evolved $\rho(\eta)$ shown in figure \ref{Fig:2} (right panel). The difference between the Lynden-Bell equilibrium and the simulated distribution is highlighted.}
\end{figure}
Since plasmas expected to relax to Lynden-Bell equilibria should be as collisionless as possible, the best testbed currently feasible for such relaxation is 1D-1V simulations, which afford the largest number of particles per cell, and, therefore, the lowest collisionality due to the reduced noise levels \cite{Birdsall_1985,Touati_2022}. As well as this, one can only expect the Lynden-Bell relaxation to be relevant when the plasma is sufficiently turbulent because inherent in the expectation of rapid maximization of entropy is the assumption of near-perfect mixing in phase space (we shall return below to the question of how perfect it really is). We therefore test our relaxation scenario for two-beam plasmas in 1D---violently unstable situations for which the saturated state is not likely to be obtained via quasilinear theory (cf.~\cite{Banik_2024})\footnote{Remarkably, it was conjectured (but not checked) already in the report of the first-ever nonlinear simulation of the two-stream instability that the equilibria eventually reached by the plasma could be explained with Lynden-Bell statistics \cite{Roberts_1967}.}. The beam instabilities that we study are described in detail in the Materials and Methods: the electron-only two-stream instability and the electron-positron two-stream instability. In the interest of simplicity, we have neglected the ion dynamics in the electron-only two-stream instability. Undoubtedly, this will prove to be an oversimplification at late times and on long length scales (missing the possibility of a plethora of ion-scale physics: see, e.g., \cite{Hutchinson_2024}), but it is the simplest possible framework to demonstrate our theory.

In figures \ref{Fig:1} and \ref{Fig:2}, we see that, at early times, $\rho(\eta)$ is well approximated by a single delta function, in line with the initial condition
\begin{equation}
\label{eqn:E10}
\rho(\eta,t=0) = 2\Delta v_{\mathrm{b}} \delta(\eta - \eta_{\mathrm{max}}),
\end{equation}
where $\Delta v_{\mathrm{b}}$ is the beam width. At these early times, the mean phase-space density, also shown in figure~\ref{Fig:2}, becomes very nearly Maxwellian. As time progresses, however, the mean phase-space density deviates from the Maxwellian equilibrium, forming an energy distribution with a power-law tail~$N(\varepsilon)\sim \varepsilon^{-2}$, while its waterbag content develops a low-$\eta$ asymptotic~$\rho(\eta)\sim\eta^{-1}$. For the measured values of~$\rho(\eta)$ and total kinetic energy, we can now treat equation [\ref{eqn:E9}] as a black box (the numerical method for this is detailed in the Materials and Methods) and compare its output to the measured mean phase-space density. Such a comparison for early and late times is shown in figure~\ref{Fig:3} for the the electron-only two-stream instability. We have done this for initial beam widths~$\Delta v_{\mathrm{b}}$ between~2\% and~50\% of the beam speed~$v_{\mathrm{b}}$ and found good agreement with [\ref{eqn:E9}] up to around~25\% of~$v_{\mathrm{b}}$, beyond which point the instability is insufficiently violent to drive strong relaxation over the times and domain sizes that were simulated. 

Quantitatively, this is the fundamental result of this paper. Viewed as a purely thermodynamic tool, the Lynden-Bell statistical mechanics correctly predicts the relaxation of the mean phase-space density in a two-stream unstable plasma. Qualitatively, however, it is possible to deduce how this thermodynamic tool operates, and why its output may be generic to many turbulent relaxing systems.

\subsection*{Universal Lynden-Bell equilibria}
We first note that the mean distribution [\ref{eqn:E9}] has a form very similar to the Fermi--Dirac distribution, owing to the analogy between the Pauli exclusion principle and the incompressibility of phase space: two fermions cannot be in the same quantum state and two level sets of phase space cannot be forced together. In a further analogy, the Lynden-Bell equilibrium can be roughly construed as the competition between two pieces of physics: the tendency of each level set to form a Maxwellian equilibrium~$\eta$ by~$\eta$ (as in the numerator of~[\ref{eqn:E9}]) and the incompressibility of phase space (expressed by the denominator of [\ref{eqn:E9}] being greater than unity).

When the incompressibility condition wins this competition (which is the classical analogue of the system being nearly degenerate), the system is very close to its ground state, known, in plasma physics, as the Gardner state \cite{Gardner_1963}. The existence of such ground states is the inevitable consequence of an exclusion principle. The `Gardner-restacked' minimum-energy counterpart to any given~$f$ is a distribution function~$f_{\mathrm{G}}(\varepsilon)$ that has~$\rho(\eta)$ identical to that of $f$, but is a monotonically decreasing function of solely the particle energy $\varepsilon = m|\v{v}|^{2}/2$. In this way, any further reduction of energy would require the distribution function to be larger at lower velocities, impossible without the compression of phase space, which is forbidden. Therefore, the ground states~$f_{\mathrm{G}}$ can be computed implicitly from---and are in one-to-one correspondence with--- the waterbag content $\rho(\eta)$:
\begin{equation}
\label{eqn:E11}
\rho(\eta) = \int\dd{\v{v}}\delta\left(f_{\mathrm{G}} - \eta \right) \odequ -2\dev{f_{G}^{-1}}{\eta}.
\end{equation}
The last equality is correct in 1D; in higher dimensions, it would involve the density of states.

This ground state, once computed, sets an important energy scale of the system: the energy of the Gardner distribution~$E_{\mathrm{G}}$. At energies much larger than this (such as in strongly turbulent systems), one should expect the effect of phase-volume exclusion to become sub-dominant to the effect that pushes each level set towards a Maxwellian---equivalent to the denominator of~[\ref{eqn:E9}] being approximately unity. Then one should anticipate the solution to have the approximate form
\begin{equation}
\label{eqn:E12}
P(\v{v},\eta > 0) \approx e^{-\beta \eta \varepsilon(\v{v})}F(\eta).
\end{equation}
The fugacity $F(\eta)$ can then immediately be deduced from [\ref{eqn:E7}]:
\begin{equation}
\label{eqn:E13}
F(\eta) \approx \rho(\eta)\left[\int \dd{\v{v}}e^{-\beta\eta\varepsilon(\v{v})} \right]^{-1} \odequ \sqrt{\frac{\beta \eta m}{2\pi}}\rho(\eta).
\end{equation}
The mean phase-space density can now be computed from~[\ref{eqn:E12}] and~[\ref{eqn:E13}] via~[\ref{eqn:E2}]: in~1D-1V,
\begin{equation}
\label{eqn:E14}
\langle f\rangle(v) \approx \int_{\eta_{\mathrm{min}}}^{\eta_{\mathrm{max}}} \dd{\eta} \sqrt{\frac{\beta \eta m}{2\pi}}\eta\rho(\eta) e^{-\beta \eta\varepsilon(v)}.
\end{equation}

The meaning of equation~[\ref{eqn:E14}] is physically transparent. Each level set relaxes to a Maxwellian distribution whose temperature is inversely proportional to its phase-space density: less dense portions behave as though they were lighter particles and more dense portions as though they were heavier. The resulting mean phase-space density is, therefore, a superposition of many Maxwellians with relative abundances set by~$\rho(\eta)$. At early times in our simulation,~$\rho(\eta)$ is approximately a delta function in~$\eta$, as in [\ref{eqn:E10}] (by design). This delta function selects from the integral in~[\ref{eqn:E14}] a single~$\eta$---so the mean phase-space density becomes an actual Maxwellian. This is precisely what is seen in figures~\ref{Fig:2} and~\ref{Fig:3}. We further note that this Maxwellianization is a \textit{fundamentally collisionless effect}: it is an entropic property of collisionless plasma that turbulence should initially want to push mono-energetic beams towards a Maxwellian equilibrium. That this is a collisionless effect is also obvious as the mean phase-space density is inhomogeneous and continues to evolve after reaching a Maxwellian (the behaviour that would be forbidden by collisional dynamics).

As the collisionless system evolves,~$\rho(\eta)$ changes form, as seen in figure~\ref{Fig:2}, developing a low-$\eta$ asymptotic for which~$\rho(\eta) \sim \eta^{-1}$. We may therefore assume that~$\rho(\eta)$ has a time-asymptotic limit
\begin{equation}
\label{eqn:E15}
\rho(\eta) \to \eta^{-1}v_{\mathrm{b}}G\left(\frac{\eta}{\eta_{\mathrm{max}}}\right),
\end{equation} 
where $G$ is a dimensionless function that only has strong dependence on~$\eta$ near the maximum phase-space density~{${\eta = \eta_{\mathrm{max}}}$}. Then~[\ref{eqn:E14}] can be written as 
\begin{equation}
\label{eqn:E16}
\langle f\rangle(v) = \frac{v_{\mathrm{b}}}{\varepsilon(v)^{3/2}}\int_{\beta\eta_{\mathrm{min}}\varepsilon(v)}^{\beta\eta_{\mathrm{max}}\varepsilon(v)}\!\!\!\!\!\!\!\!\! \dd{\bar\eta}\sqrt{\frac{m\bar{\eta}}{2\beta\pi}} G\left(\frac{\bar{\eta}}{\beta\eta_{\mathrm{max}}\varepsilon(v)}\right) e^{-\bar{\eta}},
\end{equation}
where we have changed the integration variable to~{${\bar{\eta} = \beta \eta \varepsilon(v)}$}. Therefore, for velocities such that~$\beta \eta_{\mathrm{max}}\varepsilon(v) \gg 1$ and~$\beta \eta_{\mathrm{min}}\varepsilon(v) \ll 1$, the~$\bar{\eta}$ integral will be a weak function of $v$, giving $\langle f\rangle$ a universal power-law tail. Recasting this asymptotic as a particle-energy distribution, we get
\begin{equation}
\label{eqn:E17}
N(\varepsilon)\propto \varepsilon^{-1/2}\langle f \rangle \sim\varepsilon^{-2}.
\end{equation} 

It is easy to show that, for~$\rho(\eta)\sim \eta^{-1}$, this high-energy scaling of the particle-energy distribution is true in any number of dimensions \cite{Ewart_2023}. There is a subtlety associated with what the value of~$\beta$ turns out to be and, therefore, whether neglecting degeneracy effects was ever a good approximation. This is treated in detail in~\cite{Ewart_2023}. Namely, it turns out that, while neglecting phase-space degeneracy is quantitatively incorrect, this does not invalidate the headline result~[\ref{eqn:E17}]. This occurs because the denominator of~[\ref{eqn:E9}] is substantially different from unity only at low velocities where high-density level sets crowd each other out in the phase space, whereas at larger velocities, this effect is unimportant---but this is precisely where we expect the~$\varepsilon^{-2}$ power law to emerge. 
\subsection*{Universality of waterbag content}
\begin{figure*}[t!]
\centering
\includegraphics[width=17.8cm,height=10.16cm]{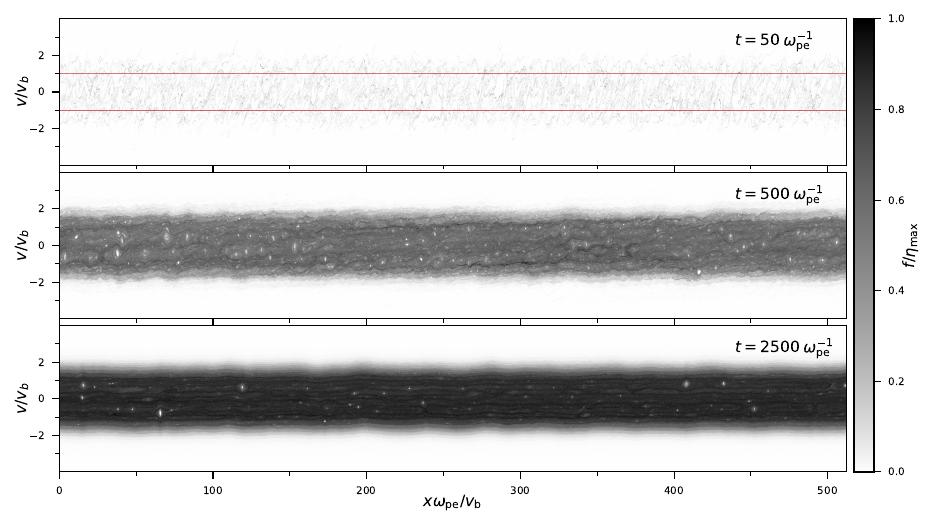}
\caption{\label{Fig:ep}Same as figure \ref{Fig:1} but for the electron-positron two-stream instability, showing the distribution of electrons (which is statistically identical to that of the positrons). While the mixing of the distribution still occurs, the formation of large-scale turbulent structures present in figure \ref{Fig:1} is entirely absent here, so the resulting distribution is much closer to its ground state. Details of the simulation setup and parameters are documented in the Materials and Methods.}
\end{figure*} 
Thus, the universal high-energy tail $N(\varepsilon) \propto \varepsilon^{-2}$ is a direct consequence of the waterbag content of the system tending towards a $\rho(\eta)\sim\eta^{-1}$ asymptotic at low $\eta$---as indeed it does in our numerical experiment. While we do not know how to prove formally that this must happen, it can be justified qualitatively in the following way. 

As follows from~[\ref{eqn:E8}], all Casimir invariants of the system can be recovered as moments of $\rho(\eta)$. In particular, the zeroth moment of~$\rho(\eta)$ is the volume of phase space in which the phase-space density $f$ takes a non-zero value. This will clearly be finite for our initial setup with two beams. As the turbulence stirs the plasma, $f$ becomes highly filamented, developing ever sharper gradients in phase space. As collisions (equivalently, particle noise \cite{Touati_2022}) smooth out these sharp gradients of~$f$, the volume in which~$f$ is non-zero should grow (this is manifest in figure~\ref{Fig:1}). However, the first moment of~$\rho(\eta)$ (which is the particle number) must stay fixed, implying that~$\rho(\eta)$ must decrease at large~$\eta$ and increase at small~$\eta$: \textit{most of the non-empty phase space must be occupied by relatively low phase-space densities}. This is manifestly (although not uniquely) satisfied by $\rho(\eta)\sim \eta^{-1}$.

Another argument to the same effect is as follows. The energy of the Gardner distribution corresponding to~$\rho(\eta)$ (the ground state defined by~[\ref{eqn:E11}]), can only increase under the action of collisional phase-space diffusion (see, e.g.,~\cite{Tremaine_1986}). It seems reasonable to conjecture, although not easy to prove rigorously, that~$f_{\mathrm{G}}$ should become more generic as it is thus heated. As was shown by~\cite{Ewart_2023}, for a wide class of~$f_{\mathrm{G}}$---all functions with any form of exponential decay at high energies\footnote{To be more mathematically precise, this means any function that does not have compact support but decays faster than any power law.}---have waterbag content with the asymptotic~{${\rho(\eta)\sim \eta^{-1}}$} as~$\eta \to 0$. 

\subsection*{Degenerate equilibria}
\begin{figure*}[t!]
\centering
\includegraphics[width=17.8cm,height=8.89cm]{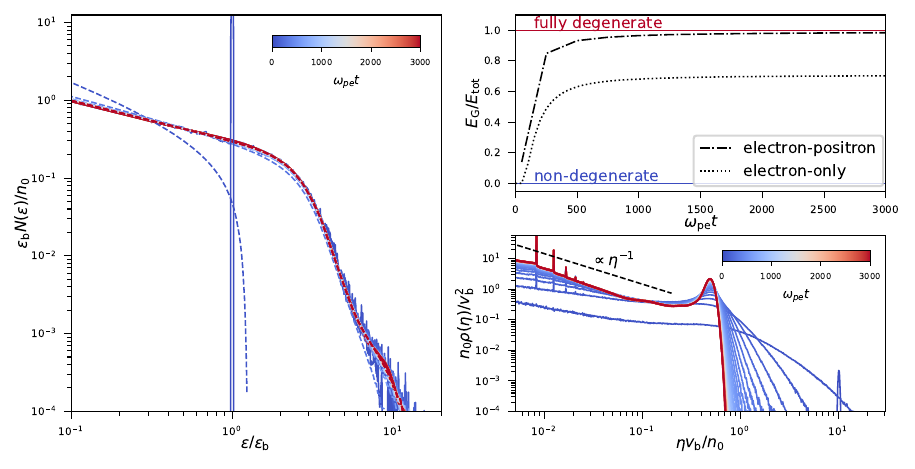}
\caption{\label{Fig:4}Left panel: comparison of the Gardner distribution function of electrons~[\ref{eqn:E11}] (dashed lines) computed from the measured waterbag content $\rho(\eta,t)$ and the mean phase-space density of electrons (solid lines) at a range of times during the evolution of the electron-positron two-stream instability. Upper right panel: ratios of the Gardner energy $E_{\mathrm{G}}$ to the total system energy $E_{\mathrm{tot}}$ during the evolution of the electron-positron (dot dashed) and electron-only (dotted) two-stream instabilities. The degenerate and non-degenerate limits are highlighted in red ($E_{\mathrm{G}} = E_{\mathrm{tot}}$) and blue ($E_{\mathrm{G}} = 0$), respectively. Lower right panel: evolution of the waterbag content $\rho(\eta)$ (of electrons) over time, as in figure~\ref{Fig:2}, but this time for the electron-positron two-stream instability. Thus, while the electron-positron system still achieves a universal waterbag content, its Gardner energy has grown so close to its actual energy that it has been frozen into its ground state.}
\end{figure*}
While the formation of the $\rho(\eta)\sim \eta^{-1}$ asymptotic may be generic, it is not the only ingredient necessary to achieve the~$\varepsilon^{-2}$ power law. This is evidenced in our second numerical experiment: the (non-relativistic) electron-positron two-stream instability---a purely numerical invention (although a reality in the relativistic setup \cite{Arrowsmith_2023,Qu_2024}), but an interesting case study because it has identical linear physics to the electron-only two-stream instability (up to a rescaling of time) but exhibits a vastly different saturation scenario. In the lower right panel of figure~\ref{Fig:4}, we see that the turbulence again pushes the waterbag content towards the asymptotic~$\rho(\eta) \sim \eta^{-1}$. However, the mean phase-space density does not form an~$\varepsilon^{-2}$ tail because the assumption of non-degeneracy [\ref{eqn:E12}] is completely violated. This can be seen in the upper right panel of figure~\ref{Fig:4}. Initially, as the beams are two thin slivers in phase space, the Gardner restacking would amount to just placing the beams at $v=0$. The resulting ground state would have an energy~$E_{\mathrm{G}}$ much smaller than the total kinetic energy~$E_{\mathrm{tot}}$, making the system highly non-degenerate. However, as the turbulence breaks~$\rho(\eta)$ conservation, the energy of the underlying Gardner distribution grows, causing the system to become more degenerate. In the case of the electron-only two-stream instability, the formation and persistence of coherent structures (phase-space holes) seen in figure~\ref{Fig:1} causes the growth of the Gardner energy to saturate. In contrast, for the electron-positron instability, the holes that do form fail to merge into large-scale, large-amplitude structures (see figure \ref{Fig:ep}). As a result, the Gardner energy increases to meet the system's total energy and the evolution freezes in a degenerate state (a phenomenon somewhat similar to `incomplete relaxation' \cite{Chavanis_2006a}). This again amounts, of course, to the system reaching its Lynden-Bell equilibrium, but in a much more trivial manner: the distribution simply becomes the Gardner distribution, as seen in the left panel of figure~\ref{Fig:4}.
\section*{Breaking of Casimir invariants and the phase-space cascade}
Let us now turn our attention to the question of why $\rho(\eta)$ is able to evolve relatively quickly (and how quickly) even in low-collisionality systems. This is because the thorough mixing in phase space, which is required for the system to reach the Lynden-Bell maximum-entropy state, generates progressively smaller scales in velocity and position space (as seen in figure~\ref{Fig:1}) until, at sufficiently small scales, collisions, however small is their rate, act to smooth the distribution function, altering $\rho(\eta)$. For electrostatic plasmas, this process has been shown theoretically \cite{Nastac_2024} and numerically \cite{Nastac_2024b} to be able to be described as a turbulent cascade through phase space of a representative Casimir invariant (cf.~{\cite{Knorr_1977,Diamond_2010,Servidio_2017,Eyink_2018}}) 
\begin{figure*}[t!]
\centering
\includegraphics[width=17.8cm,height=8.89cm]{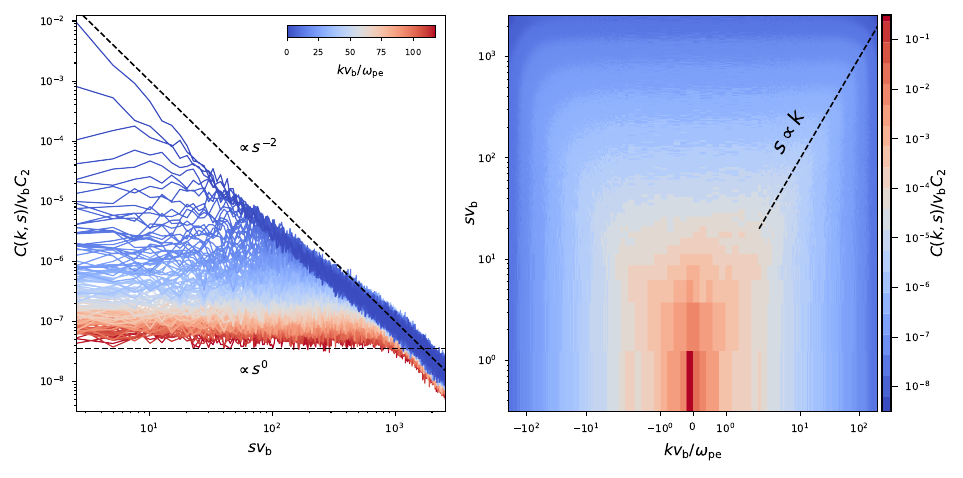}
\caption{\label{Fig:5}Left panel: cuts of $C(k,s)$ as a function of the velocity wave number~$s$~(see~[\ref{eqn:E20}]). This spectrum was taken at time~{${t = 250\,\omega_{\mathrm{pe}}^{-1}}$} during the evolution of the electron-only two-stream instability. The lines shown in the blue-to-red color palette are~$C(k,s) = |f_{k,s}|^{2}$ for different values of $k$ (plotted in increments of~$0.8\, \omega_{\mathrm{pe}}/v_{\mathrm{b}}$), representing the velocity-space structure of fluctuations of different scales (large scales: dark blue; small scales: dark red). The theoretically expected asymptotic scalings~[\ref{eqn:E21}] are shown in dashed black lines. Right panel: the same spectrum~$C(k,s)$, plotted in ~$2\mathrm{D}$ to show the so-called `critical-balance' line~$s\sim \gamma^{-1} k$, separating the two asymptotics in~[\ref{eqn:E21}]. Physically, the velocity-space linear phase mixing dominates at~$s\ll \gamma k$ and the position-space nonlinear mixing of the phase-space density by the electric field dominates at~$s\gg \gamma k$.}
\end{figure*}
\begin{equation}
\label{eqn:E19}
C_{2} = \int\dd{\eta}\eta^{2}\rho(\eta) = \frac{1}{L}\int\dd{x}\dd{v}f^{2},
\end{equation} 
where $L$ is the domain size. The distribution of this invariant across spatial and velocity scales can be quantified by its spectrum~{$C(k,s) = |f_{k,s}|^{2}$}, where
\begin{equation}
\label{eqn:E20}
f_{k,s} = \frac{1}{L}\int\dd{x}\dd{v}e^{-ikx + isv}f(x,v)
\end{equation}
is the Fourier transform of the phase-space density in position and velocity space. Since $C_{2}$ is a quadratic norm of the phase-space density, the contributions to it from the (spatial) mean~[\ref{eqn:E2}] and perturbed~$\delta f = f -\langle f\rangle $ parts of $f$ add:
\begin{equation}
C_{2} = C_{2,0} + \delta C_{\mathrm{2}} = \int \dd{v}\langle f \rangle^{2} + \frac{1}{2\pi}\sum_{k\neq 0}\int \dd{s}C(k,s).
\end{equation}
It is not hard to see that the relaxation of the initially unstable state will lead to $C_{2,0}$ decreasing and, therefore, to~$\delta C_{2}$ receiving the balance of the~$C_{2}$ density. This results in an approximately constant flux of~$C_{2}$ towards small scales (large~$k$ and~$s$). A cascade theory in the spirit of Kolmogorov \cite{Kolmogorov_1941} leads to the following asymptotic form of the spectrum~\cite{Nastac_2024b}:
\begin{equation}
\label{eqn:E21}
C_{k,s} \propto \begin{cases}  s^{-2}, & k \ll \gamma s, \\
k^{-2}, & k \gg \gamma s,
\end{cases}
\end{equation}
where $\gamma$ is a typical shearing rate in phase space set by the amplitude of the electric field. Since the electric field~{${E = -\partial \varphi / \partial x}$} is determined, via Poisson's equation
\begin{equation}
\label{eqn:E22}
\nabla^{2}\varphi = 4\pi e\left(\int \dd{v}f - n_{0}\right),
\end{equation}
by the perturbed electron density (for the experiment with static ions), the spectrum of the electric fluctuations at small spatial scales can be determined from the~$s\to 0$ asymptotic of~[\ref{eqn:E21}]:
\begin{equation}
\label{eqn:E23}
|E_{k}|^{2} = \frac{16\pi^{2} e^{2}}{k^{2}}C_{k,s\to 0} \propto k^{-4}.
\end{equation}
\begin{figure}[t!]
\centering
\includegraphics[width=8.68cm,height=8.68cm]{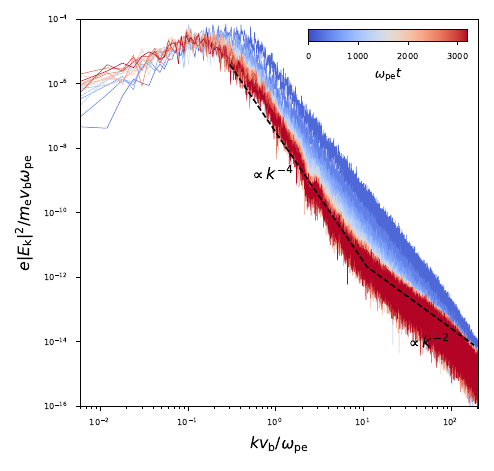}
\caption{\label{Fig:6} The spectrum of the electric field for a range of times during the evolution of the electron-only two-stream instability. The lines shown in the blue-to-red color palette represent the times from~$t = 200\,\omega_{\mathrm{pe}}^{-1}$ (dark blue) to~$t = 3200\,\omega_{\mathrm{pe}}^{-1}$ (dark red), plotted in increments of~$200\,\omega_{\mathrm{pe}}^{-1}$. The dashed black lines represent the theoretical prediction of a phase-space cascade~[\ref{eqn:E23}],~$|E_{k}|^{2}\propto k^{-4}$, which gives way to~$|E_{k}|^{2}\propto k^{-2}$, the floor due to the Poisson shot noise of discrete particles at large~$k$~(cf. \cite{Rostoker_1961,Nastac_2024b}).}
\end{figure}Because this spectrum is quite steep, the phase-space density, at any scale, is predominantly stirred by the electric field at the outer scale \cite{Batchelor_1959}, which, for the two-stream instability, are of the order of (several times) the Debye length~$\lambda_{\mathrm{D}} \sim v_{\mathrm{b}}/\omega_{\mathrm{pe}}$. It is because of this large-scale dominance that the shearing rate~$\gamma$ in~[\ref{eqn:E21}] can be assumed to be a scale-independent constant~\cite{Nastac_2024b}. This also implies that the typical time scale at which $f$ will be mixed all the way to the phase-space scale~$l_{\mathrm{c}}$ where collisions---equivalently, particle noise---start erasing the small-scale structure is \cite{Nastac_2024b} 
\begin{equation}
\label{eqn:E24}
\tau_{\mathrm{c}}\sim \gamma^{-1}\ln\frac{\lambda_{\mathrm{D}}}{l_{\mathrm{c}}}.
\end{equation}
Even without directly computing $l_{\mathrm{c}}$, it is clear that~{${l_{\mathrm{c}}\ll \lambda_{\mathrm{D}}}$}, provided the noise floor is sufficiently low (equivalent to the plasma being weakly coupled, i.e., $n_{0}\lambda_{\mathrm{D}} \gg 1$). Thus,~[\ref{eqn:E24}] tells us that~$\rho(\eta)$ will change on time scales that are only logarithmically longer than the dynamical  time scale (${\sim\gamma^{-1}}$) of collisionless relaxation. This provides a modicum of justification for our scheme---confirmed by the numerical experiment with the electron two-stream instability---of evolving the mean phase-space density as an `instantaneous' Lynden-Bell equilibrium coupled to a time-dependent waterbag content $\rho(\eta,t)$.

The above argument is supported by the excellent agreement between the theoretical predictions~[\ref{eqn:E21}] and~[\ref{eqn:E23}] and the spectra measured in our numerical simulation of the electron two-stream instability. The phase-space spectrum~$C(k,s)$ and the electric-energy spectrum $|E_{k}|^{2}$ shown in figures~\ref{Fig:5} and~\ref{Fig:6}, respectively, approach their theoretically predicted asymptotic forms around the time when the electron holes, vividly displayed in figure~\ref{Fig:1}, begin to move around and merge. It is these dynamics that provide the vigorous mixing that leads to the phase-space cascade and ultimately pushes~$\rho(\eta)$ and, therefore,~$\langle f \rangle(v)$ towards their universal forms discussed and measured above.
\section*{Discussion}
In this work, we have shown that statistical mechanics can be used to predict classes of universal equilibria for relaxing collisionless plasmas. We have tested this proposition on the example of an electrostatic plasma destabilized by one of the simplest, best-studied instabilities in plasma physics, the two-stream instability, and found good agreement with the theory.

The theory is based on the thermodynamic approach first proposed by Lynden-Bell~\cite{LyndenBell_1967}, maximising the entropy~[\ref{eqn:E3}] subject to the conservation of phase volume~[\ref{eqn:E1}]. The additional constraints arising from the latter can be captured by tracking the level sets of the phase-space density via the `waterbag content'~$\rho(\eta)$~[\ref{eqn:E7}] and endowing the resulting equilibrium~[\ref{eqn:E9}] with some memory of the plasma's earlier state. For the case of the electron two-stream instability, this theory predicts the formation of a Maxwellian distribution (despite the purely collisionless dynamics). This is indeed achieved at early times (see figures~\ref{Fig:2} and~\ref{Fig:3}). However, unlike in a collisional regime, the system continues evolving after reaching a Maxwellian. This further evolution is driven by phase-space turbulence stirred up by electron holes generated in the early stages of the instability. We show that this turbulence drives a phase-space cascade of the form predicted by~\cite{Nastac_2024,Nastac_2024b}, giving rise to small-scale structure in both velocity and position space (see figures~\ref{Fig:5} and~\ref{Fig:6}). Since no system is truly collisionless (there is always a finite number of particles), this small-scale structure causes the collisionless invariant~$\rho(\eta)$ to be broken (see figures~\ref{Fig:2} and~\ref{Fig:4}), causing the Lynden-Bell equilibrium of the system to change and the system to continue evolving on time scales that we estimate, in~[\ref{eqn:E24}], to be only logarithmically longer than the dynamical relaxation times. At these later times, the waterbag content develops a low-$\eta$ asymptotic~$\rho(\eta)\sim \eta^{-1}$, which we argue to be universal, being the natural~$\rho(\eta)$ associated with systems that have smooth ground-state phase-space densities. We show that this is indeed what happens in the numerical experiments featuring both the electron-only two-stream instability and the electron-positron two-stream instability. The details of the resulting equilibria in the presence of this universal waterbag content then depend strongly on the amount of energy the well-mixed system is able to retain relative to its ground state (defined by~[\ref{eqn:E11}]). In the case of the electron-only two-stream instability, the saturated energy of the system is larger than the energy of the ground state by a factor of order unity and, as a result, the corresponding Lynden-Bell equilibrium~[\ref{eqn:E9}] is approximately non-degenerate, featuring a distribution of particle energies that has a universal power-law tail~$\propto\varepsilon^{-2}$. In contrast, the case of electron-positron two-stream instability exemplifies systems where the non-conservation of~$\rho(\eta)$ causes the system to freeze in its ground state, achieving a fully degenerate Lynden-Bell equilibrium (cf. \cite{Hosking_2024}). 

Thus, we have two examples representing what are likely to be two equivalence classes of universal collisionless equilibria: those that, given initial energy~$E_{\mathrm{tot}}$, relax to (approximately) non-degenerate Lynden-Bell distributions such that the corresponding ground state's energy~$E_{\mathrm{G}}$ is a finite fraction of (or, better still, much smaller than) the system's energy~$E_{\mathrm{tot}}$---and those for which~$E_{\mathrm{G}}\approx E_{\mathrm{tot}}$ in the final state, which is, therefore, a fully degenerate Lynden-Bell equilibrium. It is the former class that features the universal high-energy tail~$\propto \varepsilon^{-2}$. What appears to help such a state into existence is the emergence of highly non-equilibrium structures---in the case of the electron two-stream instability, electron holes---that can store a certain amount of energy and engage in long-time nonlinear dynamics (moving around, merging) that stir up the plasma and keep it turbulent, rather than decaying into a ground state with no available energy. In a certain sense, this is similar to a non-equilibrium, driven system, where there is continuous injection of~$\delta C_{2}$ (see~[\ref{eqn:E19}]), and, presumably, other moments of~$\rho(\eta)$, into small scales---which is why the turbulent spectra that we have observed (figures \ref{Fig:5} and \ref{Fig:6}) can be predicted by theories that assume continuous driving \cite{Nastac_2024,Nastac_2024b}.

This argument contains an apparent internal contradiction: non-degenerate Lynden-Bell equilibria emerge thanks to dynamics that must clearly be inimical to the perfect satisfaction of the hypothesis of perfect mixing. Indeed, it seems unlikely that, with phase-space holes roaming the system, all parts of the phase space can be plausibly assumed equally accessible. It appears, however, that enough of it is accessible for the maximum-entropy principle to assert itself in a theoretically computable way, and that this partial accessibility is a compromise that allows the system to remain turbulent, and converge to a statistical state that is both interesting and has a modicum of universality. 

Given the propensity for distributions with power-law tails in energy to occur in a wide variety of collisionless plasmas---including, but not limited to, the solar wind (e.g.,~{\cite{Gloeckler_2008,Fisk_2014,Yang_2020}}), the solar neighbourhood (e.g.,~\cite{Oka_2018}), and numerous numerical studies (e.g., \cite{Spitkovsky_2008,Sironi_2014,Werner_2017,Zhdankin_2017})---it is interesting to speculate whether the adjusted Lynden-Bell formalism proposed (and validated) here could be used to explain some of the observed equilibria (obviously, with due regard to how the plasmas `entropic' desire to relax to Lynden-Bell states might compete with non-equilibrium processes such as particle sources and losses, energization by various structures, etc.). As has been abundantly clear since its inception, the strongest feature of the Lynden-Bell approach has also been its fundamental drawback: it is a thermodynamic theory and, while thermodynamic theories can be remarkably robust and successful, it is also difficult to predict how badly their underlying assumptions must be broken for the theory to fail completely. The idea of evolving waterbag content $\rho(\eta)$ offers some promise for understanding how the system should relax. There is, however, currently no theory of how to compute dynamically the time evolution of~$\rho(\eta)$, which is the final piece of the puzzle. Since it is this evolution that determines how far from its ground state the system saturates, it is this final piece that should teach us how to sort relaxing plasma systems between the two universality classes identified above.

\matmethods{Here, we provide some details about the set-ups of our numerical simulations of the relaxation of two-stream instabilities to Lynden-Bell equilibria. The simulations were conducted using the PIC code OSIRIS \cite{Fonseca_2002} with an initial condition containing two thin beams: 
\begin{equation}
f(\v{x},\v{v}) = \displaystyle\begin{cases}
\displaystyle\frac{n_{0}}{2 \Delta v_{\mathrm{b}}}, & v_{\mathrm{b}}-\displaystyle\frac{\Delta v_{\mathrm{b}}}{2}<|v|<v_{\mathrm{b}} + \displaystyle\frac{\Delta v_{\mathrm{b}}}{2},  \\
0, & \text{otherwise},
\end{cases}
\end{equation}
where $v_{\mathrm{b}}$ is the beams' velocity and~$\Delta v_{\mathrm{b}}$ is their width. For the simulation of the two-stream electron-positron instability, the same initial condition was used for both species. As OSIRIS is a fully relativistic code, a beam velocity of~$v_{\mathrm{b}} = c/20$ was chosen, so that the simulation was approximately classical with the relativistic gamma factor~{${\gamma_{\mathrm{rel}} - 1 \sim 10^{-3}}$}. Both the simulation of the electron-only and electron-positron two-stream instabilities used a periodic~1D domain of size~$25.6 d_{\mathrm{e}}$ (where $d_{\mathrm{e}}$ is the electron skin depth) consisting of~$2^{15}$ cells. Both linear and quadratic interpolations for particles were checked, with no discernible difference. With these parameters, the energy and momentum were conserved up to diagnostic precision for the duration of the simulation. No smoothing was used on the fields or currents. For convergence tests, the simulation of the electron-only two-stream instability was repeated on a $1\mathrm{D}$ periodic domain of size~$210d_{\mathrm{e}}$ with~$268,800$ cells, showing good convergence and much cleaner statistics. To our knowledge, this makes this the largest, longest-run, and most collisionless simulation of the electron-only two-stream instability to date. In the main text, figures~\ref{Fig:1} and~\ref{Fig:5} use the results of the smaller electron-only simulation, while figures~\ref{Fig:2},~\ref{Fig:3},~\ref{Fig:4}, and~\ref{Fig:6} use the larger simulation. Despite the fact that the peak growth rate of the two-stream instability sits at~$k \sim \sqrt{3/8}\lambda_{\mathrm{D}}^{-1} \sim \sqrt{3/8}d_{\mathrm{e}}^{-1} c/v_{\mathrm{b}}$, the domain size was chosen as a trade off between the need for resolution of sub-Debye physics and the desire for the multiple hole mergers seen in figure~\ref{Fig:1} to drive vigorous relaxation of the distribution function towards equilibrium. The number of particles per cell was chosen to be~$20,000$ (in the case of the electron-positron simulation,~$20,000$ of each species), giving the effective plasma parameter of~$n_{0}\lambda_{\mathrm{D}} \sim 10^{6}$. Naively, this would imply that the time that it would take for collisions to modify the mean distribution function in the absence of turbulence should be~$\sim 10^{6}\omega_{\mathrm{pe}}^{-1}$. As a result, our simulations should be considered collisionless as far as the evolution of~$\langle f \rangle$ is concerned, despite the fact that~$\rho(\eta)$ is modified on much shorter time scales, for the reasons explained in the main text.  

In order to compute the waterbag content~$\rho(\eta)$ of the exact phase-space density $f$, the PIC particles must be placed on a grid in~$(x,v)$ (since they are otherwise slivers with a finite extent in position space but zero width in velocity space). Therefore, we compute the fine-grained phase-space density~$f$ on evenly spaced rectangular bins with a width of 4 cells in position space and~$0.0048v_{\mathrm{b}}$ in velocity space in the interval between~$v = \pm0.25c$. The integral of the waterbag content
\begin{equation}
\Gamma(\eta) = \frac{1}{L}\int\dd{x}\dd{v}\Theta(f-\eta) = \int_{\eta}^{\infty}\dd{\eta'}\rho(\eta') 
\end{equation}
is then computed assuming a piecewise-constant $f$ on this fine-grained grid. This quantity is computed on a grid of 1000 logarithmically spaced values of $\eta$ between the maximum phase-space density $\eta_{\mathrm{max}}$ (at that time step) and $\eta_{\mathrm{min}}$ corresponding to one particle per bin. We tested the results of this scheme for different bin sizes. The chosen bin size is a compromise between the need to have small bins in order to resolve fine-grained features of the phase-space density, and the need to have bins that are sufficiently large to capture many particles. Our results were insensitive to increasing or decreasing the bin dimensions by up to a factor of 16. Due to the shape of the PIC particles in position space, it is also possible to change the lower limit for the $\eta$ grid (as a fine-grained cell may contain a fraction of a particle). The results are insensitive to changing this value by a factor of 10 in either direction.

From $\Gamma(\eta)$ and the system's energy, one knows everything required for the Lynden-Bell equilibrium to be computed numerically with correct $\beta$ and $F(\eta)$. This is done using the iterative scheme laid out in \cite{Ewart_2023}\footnote{The scheme is modified given that $\rho(\eta)$ is computed numerically, and therefore is not guaranteed to be smooth. This amounts to changing the second-order integration in $\eta$ to first-order integration.}. This iteration continues until a Lynden-Bell equilibrium is found for which the normalized root-mean-square error in $\rho(\eta)$ is below $10^{-4}$ and the error in the energy (normalized to the Gardner energy) is below $10^{-3}$.
}

\showmatmethods{} % Display the Materials and Methods section

\acknow{It is a pleasure to acknowledge encouraging discussions with Barry Ginat, Chris Hamilton, David Hosking, and Anatoly Spitkovsky (to whom we are especially grateful for suggesting an electron-positron two-beam experiment). We thank FCCN/RNCA (Portugal) for access to MareNostrum 5 (BSC, Spain) for computational work. RJE was supported by a UK EPSRC studentship; MLN by a Clarendon Scholarship; TS by FCT, Portugal (IPFN-CEEC-INST-LA3); PJB also by FCT, Portugal (Project X-maser No.2022.02230.PTDC and Grant No. UI/BD/151559/2021); the work of AAS was supported in part by grants from STFC (ST/W000903/1) and EPSRC (EP/R034737/1), as well as by the Simons Foundation via a Simons Investigator award. This research was also supported in part by the NSF grant PHY-2309135 to the Kavli Institute for Theoretical Physics (KITP) and benefited from many vigorous interactions with the participants of their programme on ``Interconnections between the Physics of Plasmas and Self-gravitating Systems''.}

\showacknow{} % Display the acknowledgments section

%\bibsplit[5]
%Use \bibsplit to split the references from the body of the text. Value "[2]" represents the number of reference in the left column (Note: Please avoid single column figures & tables on this page.)

% Bibliography
\bibliography{NLB}

\end{document}